\documentclass[twocolumn]{revtex4}
\usepackage{graphicx}
\usepackage{epsfig}
\usepackage{amsmath}
\usepackage{graphics}
\usepackage{color}
\makeatletter
\renewcommand{\theenumi}{\arabic{enumi}}

\renewcommand{\p@enumii}{\theenumi.}
\makeatother
\begin{document}
\title{Abstract Phase-space Networks Describing Reactive Dynamics}
\author{ A. Provata$^{1}$ and E. Panagakou$^{1,2}$\\}

\affiliation{
$^{1}$Department of Physical Chemistry, \\
National Center for Scientific Research ``Demokritos", GR-15310 Athens, Greece\\ \\
$^{2}$Department of Physics,
University of Athens, GR-15771 Athens, Greece
}
\date{\today}

\begin{abstract}
An abstract network approach is proposed for the description of the
dynamics in reactive processes.
The phase space of the variables (concentrations in reactive systems)
is partitioned into a finite number of segments, which constitute the
nodes of the abstract network.
Transitions between the nodes are dictated by the dynamics of the reactive
process and provide the links between the nodes. These are
weighted networks, since each link weight reflects the transition rate
between the corresponding states-nodes.
With this construction the network properties mirror the dynamics of the 
underlying process
and one can investigate the system properties by studying the corresponding abstract
 network. As a working example the Lattice Limit Cycle (LLC) model is used. Its corresponding 
abstract network is constructed and the transition matrix elements are computed via
Kinetic (Dynamic) Monte Carlo simulations. For this model it is shown that the 
degree distribution follows a power law with exponent -1, while the average
clustering coefficient $c(N)$ scales with the network size (number of nodes) $N$ as
$c(N)\sim N^{-\nu},\>\> \nu \simeq 1.46$. 
The computed exponents classify the LLC abstract reactive network into 
the scale-free networks.
This conclusion corroborates earlier investigations demonstrating the formation of
fractal spatial patterns in LLC reactive dynamics due to stochasticity and to the 
clustering of homologous species. The present construction of abstract networks
(based on the partition of the phase space) is generic and can be implemented 
with appropriate adjustments in many dynamical systems and in time series
analysis. 
\end{abstract}
\maketitle

\section{Introduction}
\label{sec:intro}

A large variety of natural, technological and social systems which require cooperation between 
many individual units operate in the form of networks. Depending on the type
of exchange between the network nodes,
two important network categories are distinguishable: 
the spatial networks and the social networks \cite{barabasi:2006,barrat:2008}.
Typical spatial networks are the large infrastructure networks, such as
the transportation network, the road-map network, the
airline and railway networks, electricity distribution network, water-pipe networks,
 etc \cite{barrat:2004,vragonic:2005}. Spatial networks are characterised
by matter exchange between their nodes.
In the same category of spatial networks various biological
networks belong, such as the neuron networks, the blood vessels networks, the
bronchial tree, the plant root network etc.
The second major category is the social networks, which include the Internet, the
Facebook, LinkedIn, Twitter, the various societies, the authors network, the actors network etc
\cite{barrat:2004,Yoon:2007}. In social networks information is shared and exchanged between 
nodes.
Both above categories
have received considerable  attention with numerous publications, including several
review papers and books in the past 15 years 
\cite{barabasi:2002,dorogovtsev:2008, barabasi:2006,barrat:2008}. 
\par A third category, which has received less 
attention is the ``state-space'' networks or the ``phase-space'' networks, which account for
systems transitting between various states and are often associated with
time series \cite{lacasa:2009,yang:2007,dong:2013}. 
For such systems we can define 
the corresponding ``abstract networks''
 whose nodes are the different states and whose links are the transitions rates
from one state to another. As such, the abstract networks are classified in the
class of weighted networks, since the links between the various states/nodes are
weighted by the transition probabilities. The current study focuses on the properties
of the abstract state network which results from a reactive system, when its continuous
phase space is segmented into a discrete number of nodes (network of states).

\par In previous studies
 abstract networks which result from the dynamics of symbol  sequences with specific
applications in DNA sequences, in motif recognition and in chaotic maps have been
considered \cite{provata:2012,latora:2010}. These networks are based on discrete state
spaces which consists of finite sequences of symbols. The nodes are identified with 
finite symbol combinations or with specific motifs, while the links are identified either
with proximity \cite{provata:2012} or with coexistence \cite{latora:2010} of the various motifs.
Unlike in the above cases, in the current study the state space (phase space)
is continuous. In reactive
dynamics where a number of species $n$ is involved, with species concentrations $x_i$, $i=1,\cdots n$,
the phase space is $n-$dimensional. Normally, the
 concentration variables are normalised (partial concentrations), 
and thus $x_i$'s are continuous variables which can take values in the range $0\le x_i \le 1$.
Since network theory is based on a finite number of nodes, the $n-$dimensional phase space
needs to be appropriately partitioned, as will be discussed in the next section (\ref{sec:net}). 
\par As working reactive model the Lattice Limit Cycle (LLC) is used. The LLC model belongs to
 the class of predator-prey systems with the additional features that a) it possesses a stable
limit cycle with dissipative global oscillations of the species concentrations at 
the Mean Filed (MF) level and b) it is lattice compatible, i.e. it can be directly 
implemented on a lattice conserving the number of lattice sites, without need to
modify its dynamics \cite{llc:2002,provata:2003,shabunin:2003}. It is implemented here
via Kinetic Monte Carlo (KMC) simulations where stochastic effects
and local interactions are taken into account. The lattice KMC realisations of this model
give rise to fractal spatial patterns which spontaneously form due to the cooperation of the
nonlinearity of the interactions and the spatial restrictions. 
\par The reason for using the LLC model as an example for the construction of  the
phase space abstract network is the complex fractal patterns which are formed during
the system's evolution and which could give rise to nontrivial transition rates among the 
nodes of the corresponding abstract network.
As it will be shown in the next sections the elements of the transition matrix have a
long range distribution and the abstract network belongs to the class of scale free networks.
\par
In the next section we propose and describe the abstract network representation of the
reactive processes. In sec. \ref{sec:llc} the general features of the LLC reactive system  are
recapitulated, both at the MF level and using KMC simulations to account for spatial and
stochastic effects.  In sec. \ref{sec:net-llc} we calculate the abstract network transition 
matrix, the degree distribution and the average clustering coefficient
which demonstrate the network's scale free character.
Finally, we recapitulate the main results in the concluding section and we 
discuss open problems.
\section{Abstract Networks of Reactive Dynamical Systems}
\label{sec:net}
The term \textit{abstract networks in reaction-diffusion systems} proposed and
developed here
should \textit{not} be confused with the classical field of ``Chemical Reaction Networks'' 
which has a long research history, mostly in Theoretical Chemistry 
\cite{martinez:2010,leenheer:2007}.
In the classical literature scientists refer to a ``network of chemical reactions'' or 
to a ``Chemical Reaction Network'' (CRN) as a finite set of reactions among a finite 
set of chemical species. In CNRs often the products of one reaction serve as reactants in 
others. CNRs find multiple applications in Biochemistry and Analytical Chemistry 
and even in Catalysis \cite{kourdis:2010,bernal:2011,coveney:2012}. The classical CNRs,
in their spatial representations, can  also
serve under certain conditions as a reaction (or reaction-diffusion) system
for the construction of the corresponding abstract network as will be described
in the sequel.
\par In the present study we consider an abstract network of nodes, each node being an
appropriately chosen segment of the state space of a reaction-diffusion system.
Thus the dynamics of the system  is mirrored on the transitions of the network
from one part of the  state space to another. To be more precise, consider a number $n$
of reactants $X_i$, $i=1,... n$ involved into a number of reactions. The reactive scheme
for the time being needs not to be explicitly written
 and it can involve any number of reactions;
even one reaction is enough for the definition of the abstract network. During the
reactive process the various species $X_i$ are represented by respective 
partial concentrations. These partial concentrations
change with time and they are denoted with small letters as
$x_i(t)$, $i=1, ... n$. Then the phase space of the system becomes an $n$-dimensional vector 
space defined as
\begin{eqnarray}
\vec X(t)=&(x_1(t),x_2(t), ... x_n(t)),\nonumber \\
               &{\rm with}\>\>\> 0\le x_i\le 1 
\label{eq001}
\end{eqnarray}

\begin{figure}[t]
\includegraphics[clip,width=0.5\textwidth,angle=0]{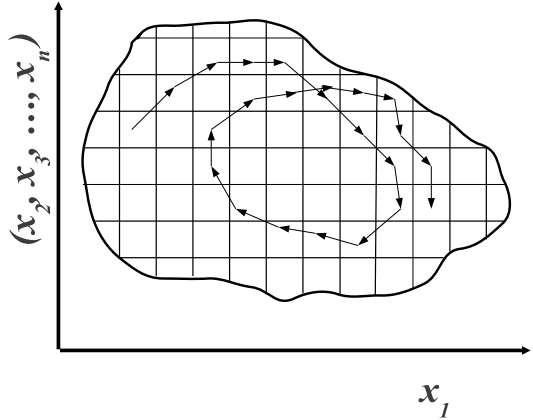}
\caption{\label{fig:01} {} Pictorial representation of the phase space
segmented into $n-$dimensional cells. Under the dynamics, 
the system's trajectory moves from one cell to another. }
\end{figure}
The state space vector depends explicitly on time $t$ and the system moves
from one point of the phase space to another with time. This motion of the
system within the phase space can be used for the creation of the abstract
network corresponding to the dynamics. Namely, we partition the phase space
into $N$ segments $S_j$, $j=1,...N$. These segments are $n-$dimensional cells,
as depicted in Fig. \ref{fig:01}. For pictorial reasons the dimension corresponding
to concentration $x_1$ is depicted on the $x-$ axis, while all other
dimensions of the phase space $(d=2, \cdots n)$ are shown collapsed on the
$y-$axis. 
\par The partition of the phase space allows us to proceed to the definition
of the abstract network characteristics. Namely, the phase space cells 
constitute the \textit{nodes} of the network, with the dynamics governing
the transitions from one node to another. An edge is drawn
between two nodes $S_i$ and
$S_j$ if the dynamics lead the system from cell $S_i$ to cell $S_j$
during one iteration step.
The edge between
$i$ and $j$ nodes is weighted with the frequency of jumps from
cell $S_i$ to cell $S_j$ during the entire system integration time.
The \textit{degree} $D_i$ of the node (cell) $S_i$ is defined 
as the fraction
of time that the system spends in space state cell $S_i$ and is equal
to the sum of weights of all links starting from cell $S_i$. In the same spirit
 the \textit{abstract network transition matrix} 
(or \textit{adjacency matrix} or \textit {connectivity} matrix) $M_{ij}$ is defined. 
The matrix element $M_{ij}$
denotes the
fraction of transitions from cell $S_i$ to cell $S_j$
during the system integration. The degree $d_j$ of node $j$ can be written
as a sum over the transition matrix elements,
\begin{eqnarray}
d_j=\sum_{i=1}^N M_{ij}, \>\>\> \{ i,j\}=1,\cdots N.
\label{eq002}
\end{eqnarray}

\par In general $M_{ij}\neq M_{ji}$ 
and thus
the abstract reactive networks belong to the class of {\it directed} networks.
Loops
(sometimes also called
''self-loops'' or ''buckles'') may be present, since the dynamics might
keep the system into the same cell after one integration step. In terms 
 of the connectivity matrix elements, the presence
of loops means $M_{ii}\neq 0$. In graph theory, graphs
which contain loops are often called {\it multigraphs}.

\par In order to study the characteristics of a network quantitative indices
are used, which 
can be computed from the transition matrix $M_{ij}$ . Based on these
 the network can be categorised in the general categories of random, small world 
or scaling network, etc. The most widely used properties  are the 
\textit{clustering coefficients} and the \textit{scaling exponents}.
The total number of nodes (cells) $N$
which cover the phase space is also called \textit{network size}.
The distribution of nodes which carry degree $d$ is denoted by $P(d)$
and describes the fractions of time the trajectory spends in the
different parts of the phase space. 
The quantity  $P(d)$ is called {\it the degree distribution}. 
It characterizes the network globally and
classifies it to be scale-free if $P(d)$
has power law tails,
\begin{eqnarray}
P(d)\sim d^{-\gamma}.
\label{eq003}
\end{eqnarray}
 $\gamma$ is the power law exponent expressing the
scale-free nature of the
network and it is typically in the range $2 <\gamma <3$,
although in some cases $\gamma$ may lie outside this interval.

\par The \textit{local clustering
coefficient} $c_l$  describes
the local network structure around the specific node  $S_l$ and
is defined as:
\begin{eqnarray}
c_l=\frac{\sum_{i,j}M_{li}M_{ij}M_{jl}}{\sum_{i \not=  j}M_{li}M_{jl}},
\>\>\> \{ i,j\}=1,\cdots N.
\label{eq004}
\end{eqnarray}
In Eq. \eqref{eq004} the numerator expresses the total
weighted number of closed triangles
originating from node $l$, while the denominator
gives the maximum number of possible
triangles originating on the same node \cite{grindrod:2002,saramaki:2007}.
\par The distribution of clustering coefficients $P(c)$
defines another critical exponent $\beta$, which
takes a power law form in scale free networks,
\begin{eqnarray}
P(c)\sim c^{-\beta}.
\label{eq0051}
\end{eqnarray}
Watts and Strogatz  have shown that $ P(c)$ is exponentially decaying for random,
uncorrelated networks \cite{watts:1998,ravasz:2003}.

\par The global clustering coefficient $c(N)$, defined as
the average of the local clustering
ones, characterises globally the connectivity in the network
and in general depends on the size $N$ of the network,
\begin{eqnarray}
c(N)=<c_l>=\frac{1}{N}\sum_{l=1}^N c_l.
\label{eq005}
\end{eqnarray}

$c(N)$ is important in the characterisation of typical networks. 
In particular, in  random uncorrelated
networks $c(N)$ decreases as \cite{watts:1998}
\begin{eqnarray}
c(N)\sim N^{-1}.
\label{eq006}
\end{eqnarray}
In the case of scale-free, highly clustered and complex
networks Eq. \eqref{eq006} takes the general form
\begin{eqnarray}
c(N)\sim N^{-\nu}.
\label{eq007}
\end{eqnarray}

\par It is worth here to understand
 how the phase space of the system behaves under different dynamical
regimes and the form that the corresponding transition matrix takes. 
Indicative cases are:
\begin{enumerate}
\item When the phase state of the system contains a single attracting fixed
point, then as $t\to \infty$ all initial states will converge to this attractor.
\begin{eqnarray}
\lim_{t\to\infty}\vec X(t)&=&\vec X_{fix}\nonumber\\
&=&(x_{1,f}(t),x_{2,f}(t), ... x_{n,f}(t))
\label{eq008}
\end{eqnarray}
In this case the transition matrix will only contain one single non-zero element,
namely, the diagonal element which corresponds to the cell containing the
attracting fixed point.
\item When the phase space of the system contains multiple attracting fixed
points, then as $t\to \infty$ each initial state will converge to one of 
these attracting fixed points depending on the position of the state space
vector at $t=0$. In this case the transition matrix will contain several
scattered non-zero elements at the positions which correspond
 to the cells containing the attracting fixed points.
\item When the phase space of the system contains attracting periodic orbits
(deterministic limit cycle)
the state space vector will soon enter in one of those and will continue to
move therein, following the orbit. Then only a subpart of the phase space is
visited in the long run. The same is true for conservative systems
where  an infinite number of periodic orbits exist.
These specific closed trajectories will be mirrored in the form of
the transition matrix.
\item When the phase space of the system contains a chaotic attractor
the state space vector eventually enters the attractor, is trapped therein 
and continues to move only in the area covered by the attractor. In this
case also, the structure of the transition matrix and the critical
exponents characterise the 
dynamics of the chaotic trajectory.
\end{enumerate}
It must be noted here that in order to obtain numerically the 
correct transition matrix one should 
use very long integration times of the system so that the initial states 
will be forgotten and their overall contribution to the transition matrix
will be negligible.
\par Although in the former two cases (1 and 2)
the structure of the transition matrix is
somewhat trivial, the same is not true in the latter two cases. Especially, the
case of chaotic attractors should lead to nontrivial transition matrices 
and critical exponents, and its structure
could be compatible with small world or scale free networks.
\par In the next section the particular reactive dynamics is specified as the Lattice
Limit Cycle dynamics and its network characteristics are computed
via KMC simulations which take into account spatial restrictions
and stochastic effects.

\section{Working Example: The Lattice Limit Cycle Model}
\label{sec:llc}

The LLC model has been previously studied and shown to possess many interesting properties:
a) it has strong non-linearities of 4-th order, b) it is lattice compatible, in the sense 
that it can be directly applied on a lattice, without modifications on its dynamics, c) 
implies single-particle occupancy of a lattice site with the possibility of keeping certain 
sites empty d) both at the MF level and in the simulations it leads to dissipative oscillations
e) gives rise to spatiotemporal pattern formations and fractal structures and f) gives
rise to synchronization phenomena \cite{llc:2002,provata:2003,shabunin:2003,tsekouras:2006}.
Due to its complex but tractable dynamics the LLC is an ideal candidate model for testing the 
abstract network approach and
for studying the various reactive properties in view of the network characteristics.
\par In the next three subsections we recapitulate the most important features of the 
LLC model which will be used in the sequel for the study of the reactive abstract network properties.
\subsection{The Model}
\label{sec:llc-model}

The LLC model involves two types of interacting species (or particles) $X_1,X_2$ and one
virtual species $S$. The two interacting species  participate in a series of reactions 
on an underlying surface, while the virtual species $S$ represents the empty surface sites. 
Without loss of generality, the surface is assumed to have
 the form of a square lattice of size $L \times L$, containing $L^2$ reactive sites. Each site 
can host at most one particle ( $X_1-state$ or $X_2-state$),
 or it can be empty ($S-state$). The reactive scheme is the following:

\begin{subequations}
\begin{equation}\label{eq01a}
2X_1 + 2X_2  \stackrel{p_1}{\rightarrow}  3X_2+S
\end{equation}
\begin{equation}\label{eq01b}
S + X_1  \stackrel{p_2}{\rightarrow}  2X_1
\end{equation}
\begin{equation}\label{eq01c}
X_2 + S  \stackrel{p_3}{\rightarrow}  2S
\end{equation}
\label{eq01}
\end{subequations}

Each of the three reactions takes place with corresponding rate $p_i$; all rates will
be translated into probabilities in the simulation process. The meaning of
scheme \eqref{eq01} is the following:
 when two particles of type $X_1$ are found within reaction distance
with two particles of type $X_2$ they react with rate (probability) $p_1$ and 
one of the $X_1$ turns into $X_2$ while the other $X_1$
turns into $S$ (desorbs or dies). 
Likewise, when an empty site $S$ is found within reaction distance with a
particle $X_1$ it reacts with probability $p_2$ and
the empty site acquires an $X_1$ particle (adsorption). When an
$X_2$ particle is found within reaction distance with an empty site $S$ it reacts desorbing
 with probability $p_3$. Reactive steps \eqref{eq01b} and \eqref{eq01c} are known
in literature as cooperative birth and death processes, respectively. 
\par The LLC scheme represents a non-equilibrium physicochemical process, with adsorption
and desorption (or birth-and-death) mechanisms. The system is then considered as open and 
particles (species) can enter or leave the lattice. In some cases the particles can be immobile on the
lattice and only reactions between particles are allowed within a certain distance. A
common example is the disease spreading in (immobile) plants \cite{sokolov:2007}. In other cases
some of the particles maybe mobile and diffuse on the lattice. A common example is the
$CO$ oxidation on the surface of $Pt$. In this process $CO$ molecules and atomic oxygen
$O$ are adsorbed on the surface site of $Pt$. The $CO$ molecules are highly mobile and
diffuse on the lattice, while the $O$ atoms are strongly attached on the surface and 
they do not move \cite{pavlenko:2003}.
\par
It is important to stress here that the realisation of step \eqref{eq01a} is difficult
on lattice due to the requirement of simultaneous presence of the four specific particles, 
(two $X_1$ and two $X_2$) in the same local neighbourhood. And although at the MF
level the dynamics of step \eqref{eq01a} is governed solely by the kinetic constant $p_1$, in
the lattice realisation the local geometry and connectivity are shown to play an equally crucial role 
\cite{llc:2002,tsekouras:2006}.
In particular, the dimensionality and type of lattice (square, triangular, cubic etc) is important in the
sense that the critical points, the equilibrium states, the fixed points and the amplitude
of oscillations depend quantitatively on the lattice type and dimensionality 
\cite{zhdanov:1999,provata:2005,zhdanov:1992}.
However, the qualitative features are common in most lattice types, with the exception 
of the 1-D lattice, where the reaction scheme \eqref{eq01a} is difficult to be realised due
to the restricted local geometry (only two first-neighbour sites for each node of the lattice).
That is why, without loss of generality, we have decided to use the square lattice configuration 
as a substrate and to investigate the properties of the corresponding reactive
abstract network.

\subsection{Mean Field Equations}
\label{sec:llc-mf}
Deterministic MF is the traditional approach to reactive dynamics and although it does not describe 
the processes in detail it gives the general properties and tendencies of the phase
space of the system. The MF equations describing the kinetic scheme \eqref{eq01} are
\cite{llc:2002}:

\begin{subequations} \label{eq02}
	\begin{equation}	\label{eq02a}
	\frac{dx_1}{dt}  =  -2p_1 x_1^2 x_2^2  +  p_2 s x_1
	\end{equation}
	\begin{equation}	\label{eq02b}
	\frac{dx_2}{dt}  =  p_1 x_1^2 x_2^2  -   p_3 x_2 s 
	\end{equation}
	\begin{equation}	\label{eq02c}
	\frac{ds}{dt}  =  p_1 x_1^2 x_2^2 -p_2 x_1 s+  p_3 s x_2
	\end{equation}	
\end{subequations}
\noindent
where the small letters $x_1, \> x_2$ and $s$ denote the temporal 
MF concentrations of the particles $X_1, \> X_2$ and $S$, respectively. It is an 
inherent property of the system to satisfy the space conservation condition, i.e.
that the total number of $X_1$, $X_2$ particles plus the empty sites $S$ should be
conserved. This is also mirrored in the MF conservation condition $x_1+x_2+s=constant$
which is directly satisfied by the MF Eqs. \eqref{eq02}. Customarily, the
constant is identified as unity and then the variables $x_1,\> x_2 ,\>s $ are identified with the 
partial concentrations of the corresponding species. In a further simplification the
system \eqref{eq02} is reduced  to two equations,
by eliminating the variable $s$ via the conservation condition $s=1-x_1-x_2$ \cite{llc:2002}.
The reduced nonlinear system, also of 4-th order, reads:
\begin{eqnarray}
\frac{dx_1}{dt}&=&-2p_1x_1^2x_2^2+p_2x_1(1-x_1-x_2)
\nonumber \\
\frac{dx_2}{dt}&=&p_1x_1^2x_2^2-p_3x_2(1-x_1-x_2)
\label{eq03}
\end{eqnarray}

Both original and reduced LLC have 4 fixed points $Q_i$, three of which are trivial. In the
reduced representation the fixed points are: 

\begin{subequations}
\begin{equation}\label{eq04a}
Q_1 = (0,0)  
\end{equation}
\begin{equation}\label{eq04b}
Q_2 = (0,1) 
\end{equation}
\begin{equation}\label{eq04c}
Q_3 = (1,0)  
\end{equation}
\begin{eqnarray}\label{eq04d}
Q_4=\left(  \sqrt[3]{\frac{p_3^2}{p_1p_2}\left[1+K\right]}
+\sqrt[3]{\frac{p_3^2}{p_1p_2}\left[1- K\right]}, \right. \nonumber \\
\left.  \sqrt[3]{\frac{p_2^2}{8p_1p_3}\left[1+ K\right]}
+\sqrt[3]{\frac{p_2^2}{8p_1p_3}\left[1- K\right]} \right) 
\end{eqnarray}
\label{eq04}
\end{subequations}

\noindent with constant $K=\sqrt{1+(2p_3+p_2)^3/(27p_1p_2p_3)}$.
While $Q_i$, $i=1,2,3$ are saddle fixed points, the stability of $Q_4$ 
is nontrivial. Depending on the parameter values,
$Q_4$ undergoes a supercritical
Hopf bifurcation giving rise to a limit cycle and periodic oscillations
of the species concentrations.
\par The analysis so far refers solely to the deterministic MF description of the LLC,
without taking into account 
spatial restrictions and stochastic effects, which will be briefly
recapitulated in the next subsection. 

\subsection{Classical Kinetic Monte Carlo Simulations}
\label{sec:llc-kmc}

As stated earlier in this study and in previous publications 
\cite{nicolis:1977,zhdanov:1999,llv:1999}
the MF theory describes reactive systems only in the case of well mixed systems when
every particle can interact equally with all other particles in the system, independently
of their distance and position. Thus MF conditions are only achievable in well-stirred systems
and they do not hold in systems where local effects need to be taken into account
 \cite{nicolis:1977}. For systems with spatial restrictions or with local stochastic
effects the most detailed method is the probabilistic Master Equation
approach, which accounts for all possible transitions of the system from
one state to another, taking into account all spatial and stochastic effects.
When the number of states is small, the probabilistic Master Equation is very
efficient but it becomes intractable for large systems with many details in their
spatial architecture. KMC methods are customarily designed to overcome this difficulty.
They advance the system dynamics based on probabilistic transitions taking place
by comparing the reaction rates with random numbers. The KMC method used here to
describe the LLC model assumes single occupancy of every lattice site and 
interactions between nearest neighbouring sites. The rates are transformed into
probabilities by dividing each one of them with the sum of the others.
This transformation introduces a rescaling in time, without affecting the qualitative
features of the system.
Without loss of generality, the lattice here is assumed to 
be square with cyclic boundary conditions. The algorithm is described 
by the following steps:

\begin{enumerate}
\item A random site $(i,j)$ on the lattice is chosen.\label{step1}
\item If the site $(i,j)$ contains an $X_1$ particle and among its first neigbhours
one $X_1$ particle and two $X_2$ particles are found, then the chosen $X_1$ particle
and the neighbour $X_1$ particle are changed to $X_2$ and $S$, in random order. This step
represents reaction scheme \eqref{eq01a} and takes place with probability $p_1$.
\item If the chosen site $(i,j)$ contains an $S$ particle (empty site) 
and a neighbouring site 
contains an $X_1$ particle, then the chosen 
site $(i,j)$ adsorbs an $X_1$ particle ($S$ is replaced
by $X_1$) with probability $p_2$. This step is a cooperative adsorption or birth event and 
represents reaction scheme \eqref{eq01b}.
\item If the chosen site $(i,j)$ contains an $X_2$ particle  and a neighbouring site 
contains an $S$ particle
 (empty site), then the chosen $X_2$ desorbs ($S$ is replaced
by $X_2$) with probability $p_3$. This step is a cooperative desorption or death event and 
represents reaction scheme \eqref{eq01c}.
\item In all other cases the lattice remains unchanged.
\item An Elementary Time Step (ETS)
 is completed and the algorithm returns to step \eqref{step1} for a new reaction ETS to start.
\end{enumerate}

\par
The time unit, Monte Carlo Step (MCS) of this process is defined as the number of ETS
 equal to the total number  of lattice sites, i.e. for a square lattice of
linear size $L$, $1\>\> MCS=L^2 \> ETS$. With this definition each lattice site has the chance
to react once, on average, in each MCS \cite{llc:2002} while at each ETS at most one
reactive event takes place.
\par
\begin{figure}[t]
\centering
\includegraphics[width=0.5\textwidth,angle=0]{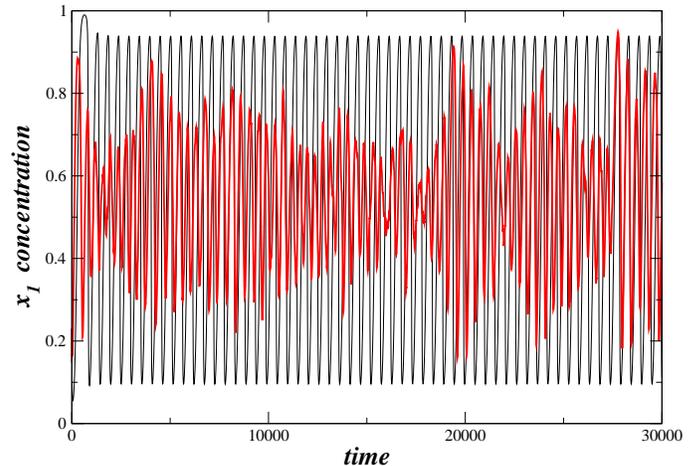}
\caption{\label{fig:02} {}
Temporal evolution of the $x_1$ concentration in LLC for parameter values 
$p_1=0.9585,p_2=0.016, p_3=0.026$ and lattice size $L=2^7$.
The solid black line depicts the MF integration with purely periodic cycles,
 while the thick red line denotes the
KMC simulations with irregular oscillations. }
\end{figure}

 As shown in previous studies, while the MF dynamics dictates the presence of a 
limit cycle, after the system passes through a Hopf bifurcation, the reduction of the
system on a low dimensional support together with the stochastic noise induced on the
 dynamics, lead to intermittent oscillations of the species concentrations on the
lattice \cite{llc:2002,provata:2003,tsekouras:2006}. In Fig. \ref{fig:02} we present
the temporal evolution of the LLC for the working parameter set $p_1=0.9585,p_2=0.016, p_3=0.026$, 
which lies beyond the Hopf bifurcation point. The black solid line denotes the MF dynamics:
after an initial transitory time, the system generates regular cycles whose amplitude
depends solely on the parameters $p_i$. The thick red line is generated by KMC simulations
for the same parameter values and for system size $L\times L=2^7\times 2^7$. 
The periods (and frequencies) of the two time series
are very close, but the KMC oscillations are irregular and show intermittent bursts.
As previously discussed \cite{llc:2002,tsekouras:2006} the spatial extension of the
system together with the limited range of the interactions divide the system into
local oscillators which operate out of phase. Intermittent oscillations are observed
for finite system sizes due to synchronisation of clusters of oscillators, while for
large system sizes oscillations are suppressed due to cancellation of the random
phases.

Having briefly recapitulated the detailed dynamics and spatiotemporal evolution of the LLC system
we next examine the representation of this dynamics in terms of the corresponding
abstract reactive network.  

\section{Network features of the Lattice Limit Cycle Model}
\label{sec:net-llc}

For the realisation of the network it is possible to use an $n=1$ phase space representation, where only the
$X_1$ (or only the $X_2$) species is considered or an $n=2$ representation where both species
$X_1$ and $X_2$ are simultaneously recorded. We first give examples to show that many 
network features are the same whether we use the $n=1$ or the $n=2$ representation of the system.

\par Starting with the $n=1$ phase space, $X_1$ taken as an example, the phase space is 
partitioned into $N_1$ segments $S_i$, $i=1, \cdots N_1$ of size $\epsilon_1$, where
\begin{eqnarray}
S_i=\left[ x_1^{min}+i\epsilon_1 , x_1^{min}+(i+1)\epsilon_1 \right]
\label{eq11}
\end{eqnarray}
These segments $S_i$ are identified as the $N_1$ discrete states of the system.
In Eq. \eqref{eq11} $x_1^{min}$ is the minimum value of the variable $x_1$ and $x_1^{min}+
N_1\epsilon_1=x_1^{max}$. Next, the transitions rates from segment (state) $S_i\to S_j$ 
are recorded
during the KMC simulations in each Monte Carlo step. 
Based on these rates the transition matrix $M_{i,j}$ of size $N_1\times N_1$
can be written.  This is the 
matrix that characterises the abstract network, its adjacency matrix.
Since the domain of $x_1$ values is equipartinioned, 
some of the segments might not be visited,
due to the dynamics, especially in the case where the $x_1-$domain is not fully
connected. In this
case some rows in the adjacency matrix will have all their elements equal to zero.
Independently, the same construction can be designed for the variable $x_2$, whose
phase space is segmented into $N_2$ nodes, while the corresponding adjacency matrix has size 
$N_2 \times N_2$. Depending on the particular problem $N_1$ can be different from, or 
equal to $N_2$. 
\begin{figure}[h]
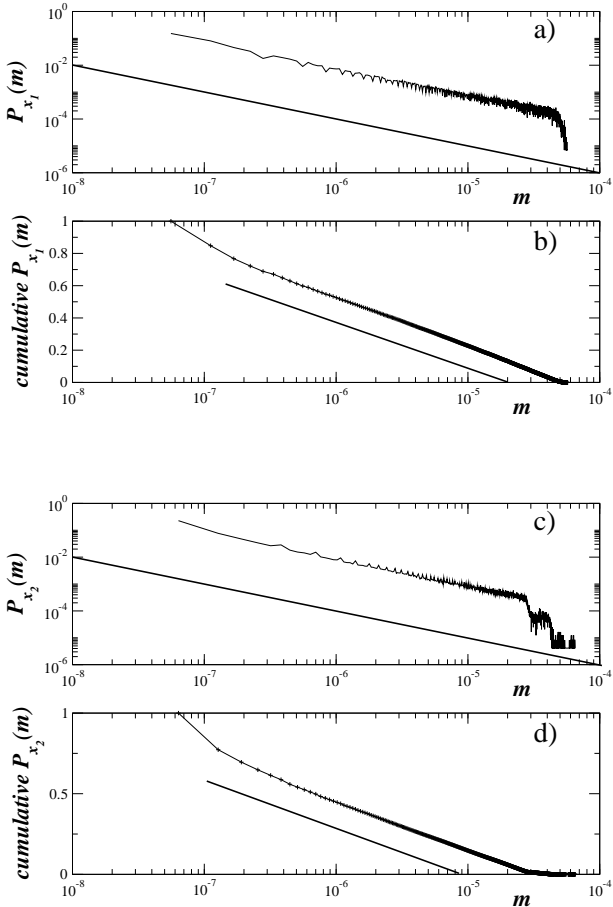

\includegraphics[width=0.45\textwidth,angle=0]{./fig03ab.eps}
\vskip 1cm
\includegraphics[width=0.45\textwidth,angle=0]{./fig03cd.eps}
\caption{\label{fig:03}{}
The size distribution of the elements of the transition matrix $M_{i,j}$
in the $n=1$ phase space representations: a) Size distribution of elements related
to the variable $x_1$. The
solid straight line indicates a power power law with exponent -1. b) 
Cumulative size distribution of elements related
to the variable $x_1$. The solid line indicates a logarithmic dependence.
c) Size distribution of elements related
to the variable $x_2$. The
solid straight line indicates a power power law with exponent -1. b) 
Cumulative size distribution of elements related
to the variable $x_2$. The solid line indicates a logarithmic dependence.
Parameter values are
$p_1=0.9585,p_2=0.016, p_3=0.026$.
The system size is $L\times L=2^8\times 2^8$ and $T=5\cdot 10^7$ Monte Carlo
iteration steps were computed. }
\end{figure}

 \par The size distribution of the $M_{i,j}$ elements (transition probabilities)
 is a very important statistical property of the network and can characterise its type.
In Fig. \ref{fig:03}a we plot the size distributions $P_{x_1}(m)$ which represents
the probability that
that a matrix element $M_{i,j}$
is in the interval $\left[ m, m+dm\right] $ in the $n=1$ phase space representation. 
The parameter values are chosen as $p_1=0.9585,p_2=0.016, p_3=0.026$ and they are
located in the region where intermittent oscillations are observed in the KMC simulations
(see Fig. \ref{fig:02}).
In a double logarithmic scale the size distribution roughly follows a power law decay
$P_{x_1}(m)\sim m^{-\gamma_1}$ with exponent $\gamma_1=1$. The value of the exponent
$\gamma_1$ is further verified from the cumulative distribution shown in Fig. \ref{fig:03}b.
In a single logarithmic scale ($x-$axis) the cumulative distribution takes a linear
form, indicating logarithmic dependence which further points to a power law exponent
of the order of $\gamma_1 =1$ for the original distribution. Similar behaviour is also recorded
for the adjacency matrix corresponding to the $x_2$ variable. In Fig. \ref{fig:03}c the $P_{x_2}(m)$ 
distribution is plotted in a double logarithmic scales. Similarly to Fig. \ref{fig:03}a
a scaling $\sim m^{-1}$ is followed,
while in Fig. \ref{fig:03}d the cumulative $P_{x_2}(m)$ distribution shows logarithmic dependence.
This similar scaling is expected, since the two variables interact feeding each other
and describe the dynamics of the same system.

\par
\begin{figure}[t]
\includegraphics[width=0.45\textwidth,angle=0]{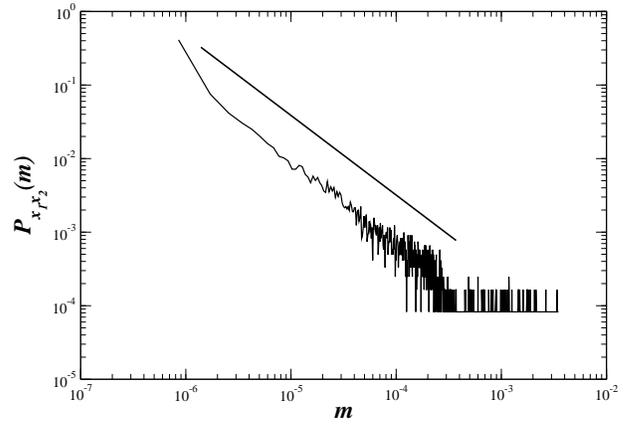}
\caption{\label{fig:04}{}
The size distribution of the elements of the composite transition matrix $M_{i,j}$
in the $n=2$ phase space representation. The straight solid line has slope -1.08.
Parameter values and system size as in Fig. \ref{fig:03}. }
\end{figure}

 The composite, $n=2$, network representation requires longer matrices since the phase 
space contains $N_1 \times N_2$ nodes. Then the adjacency matrix has dimensions
$N_1N_2\times N_1N_2$. In Fig. \ref{fig:04} the size distribution $P_{x_1x_2}$
of the $n=2$ adjacency matrix is plotted, for the KMC simulations using the working 
parameter set. The phase space is segmented in $64^2\times 64^2$
cells, using $N_1=N_2=64$.  
For comparison, the solid straight line in the double logarithmic 
scale represents a power law with exponent $\gamma_1=1.08$, which is consistent with
the one observed in the $n=1$ representations. This nontrivial nonexponential decay
of the size distribution of the transition matrix elements indicates that
the abstract network which was constructed from the reactive process is a
scale free network.
\par
\begin{figure}[t]
\includegraphics[width=0.45\textwidth,angle=0]{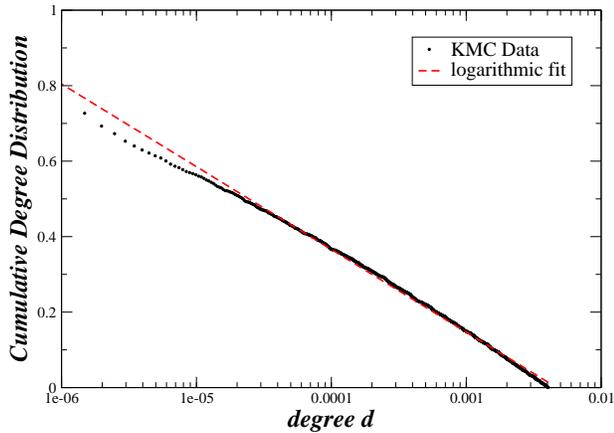}
\caption{\label{fig:05}{}
The degree distribution of the network nodes
in the $n=2$ phase space representations.
Parameter values and system size as in Fig. \ref{fig:03}. }
\end{figure}

The degree distribution of the network nodes presents similar power law features.
For the calculation of the degree distribution the $n=2$ representation is
used, which accounts for the phase space properties in more detail. Because
the degree $d_j$ of a specific node $j$ is composed as a sum over 
the transition matrix elements $M_{ij}$  (see Eq. \eqref{eq002})
and the size distribution of the
matrix elements  follows a power law then the degree distribution follows the same
power law  provided that the number of nodes $N$ is large enough. This is a direct
consequence of the Generalised Central Limit Theorem which addresses
distributions with power law tails. Indeed, in Fig. \ref{fig:05} the 
cumulative degree distribution of the network nodes is depicted and shows clear
linear behaviour in semi-logarithmic scale. This is consistent with a power law
degree distribution with exponent $\gamma =1$, as was also observed for the size
 distribution of the transition matrix elements. 
Note that if the matrix elements
$M_{ij}$ are randomly and uniformly distributed then 
 the degree distribution takes the shape of a noisy Gaussian (for finite size
transition matrices).
\par Small degree distribution exponents,
 $\gamma <2$,  are not unusual in real world phenomena; they have been observed
and studied in particular in social networks (e-mail networks, package exchange
networks etc.). It expresses the property of some networks
 where the total number of links or total weight carried by links 
grows faster than the number of nodes \cite{seyed:2006}.

\begin{figure}[t]
\includegraphics[width=0.45\textwidth,angle=0]{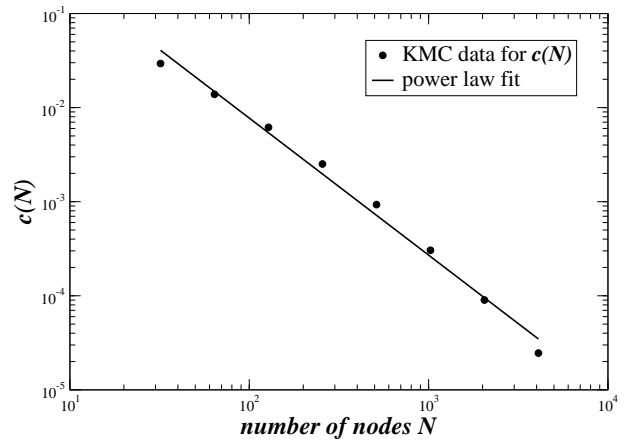}
\caption{\label{fig:06}{}
The average clustering coefficient as a function of the network size.
Parameter values and system size as in Fig. \ref{fig:03}. }
\end{figure}

The distribution of clustering coefficients $P_{cc}$ also presents power law decay
while the distribution $P_{cc}$ in the case of  random and uniformly distributed
transition matrix elements (links) $M_{ij}$ acquires a (noisy) Gaussian form.
To further probe on the scale free properties of the abstract network
we calculate the global (average) clustering coefficient as a function of the network
size which is represented by the number of network nodes $N$. For simplicity we
study  the phase spaces of the $x_1$ 
and $x_2$ variables separately, to keep the size of the transition
matrix to order of $N_{1(2)}\times N_{1(2)}$, for computational convenience. The phase spaces of the
variables $x_1$ and $x_2$ are individually 
segmented in $N=32, \cdots 4096$ nodes and the corresponding
average clustering coefficients $c_1(N)$ and $c_2(N)$ are computed. 
 Figure \ref{fig:06} depicts the global clustering coefficient $c(N)=\left( c_1(N)+c_2(N)
\right) /2$.  In a
double logarithmic scale $c(N)$  shows a power law decay with exponent $\nu \sim 1.46$.
Similar values of the exponents are calculated for the individual clustering coefficients
based on the $x_1$ and $x_2$ concentrations. Note that the exponent $\nu =1$ 
characterises random and uncorrelated networks \cite{watts:1998},
while the value $\nu =1.46$ computed here for the LLC model further verifies that this system is
in the category  of scale free networks.
 
\par The presence of power law exponents in the network representation
of the reactive system KMC realisation on low dimensional lattice supports
corroborates earlier investigations which demonstrate the formation
of fractal patterns in the spatial arrangement of the reactants and products
as a result of clustering of homologous particles and of cluster-cluster
competition \cite{tsekouras:2006,sokolov:2007}. 
As stochastic interactions take place at the borderlines
between different clusters these borderlines acquire complex fractal patterns
and this spatial complexity, enhanced by the activity dynamics, shapes the form of the
transition matrix. Thus the network power law exponents have their origin in the
fractal spatial patterns developed as a result of the support 
restrictions on the dynamics of the reactive process.

\section{Conclusions}

\par In the current study an abstract network construction is proposed for reactive systems
based on the coarse graining of their phase space. The phase space of the reactive system,
which is composed by the partial species concentrations, 
is divided into a numbers of segments which serve as the nodes of the abstract network.
As described in sections \ref{sec:net} and \ref{sec:net-llc},
the trajectories which represent the temporal evolutions of the partial concentrations direct
the system from one node to the next, determining the
jump frequency between the different nodes and thus 
providing the adjacency matrix. Having the adjacency matrix it is then possible to compute
all the network properties such as degree distribution, clustering coefficients, critical
exponents and to classify the network as scale free, Erdos-Renyi, Small World etc.
\par As working example the Lattice Limit Cycle model is used which is an open, far from 
equilibrium, nonlinear, reactive system involving three reactive species participating in
 autocatalytic reactions, cooperative desorption and adsorption processes. The system
is restricted on a two-dimensional lattice support with single particle lattice occupancy
and local nearest neighbour interactions. The temporal evolution proceeds
 via KMC simulations which introduce stochastic factors in the system's integration. 
Although the deterministic dynamics at the MF level predict dissipative oscillations of
limit cycle type, the stochastic effects induced by the KMC dynamics lead to intermittent
oscillations (with average amplitude depending on the system size) and the formation of
spatiotemporal fractal patterns. 
\par It is shown that the abstract phase space network corresponding to this reactive
system is scale free and is characterised by a power law degree distribution $P(k)\sim k^{-\gamma}$ with 
exponent $\gamma \simeq 1$ while the average clustering coefficient $c(N)$ increases with the
network size $N$ as $c(N)\sim N^{-\nu}, \>\> \nu\simeq 1.46$. The origin of this  nontrivial 
clustering is attributed to the cooperative nature of the reactions which form fractal
spatiotemporal patterns on the lattice. The cooperative effect requires the presence of
two different particles in adjacent lattice sites in order to react and this gives rise
to segregation and to formation of clusters of homologous species. As interactions take
place stochastically only in the borders between the clusters, surface roughening takes
place producing fractal borderlines. These spontaneously produced fractals are at the
 origin of the scale free nature of the  corresponding abstract phase space network. 

\par The same general idea of abstract phase space networks can be applied to all
reactive systems as well as in reaction-diffusion systems. The properties of the
corresponding network reflect the spatial distribution of the reactants and together
with the dynamics (nonlinearities) of the interactions determine the qualitative
and quantitative properties of the network.

\par The construction of the abstract phase space network was based on the discrete time series
which resulted by appropriately coarse graining the phase space of the concentrations variables
($x_1(t)$ or/and $x_2(t)$) and by following the dynamics as the system's trajectory crosses from one cell
to another in time. The discrete time series was the only information used for the subsequent
construction of the abstract network. Equivalently, if a time series 
$f_i, \>\>, i=1,...T$ is 
given, $f$ can be considered as the trajectory of an unknown dynamical system. 
 Network analysis of time series has been the subject of various studies using 
different constructions of the phase space to study fractional Brownian motion 
and other processes \cite{lacasa:2009,yang:2007,dong:2013}. In the current LLC
study the weighted network construction is based on the calculation of the
total number of transitions between nodes and not on the local distances of the nodes
on the reconstructed phase space.
This way  the adjacency matrix elements
are calculated as the trajectory travels from node to node in the phase space,
provided that the coarse grained time series of the concentrations $x_i, \>>
i=1, \cdots T$ is long enough
to be considered as stationary (to have escaped the transient regime).
This way any time series which has attained stationarity can be described by an abstract phase
space network and the network characteristic exponents mirror the correlations of the 
time series. 
It would be interesting to investigate the network characteristics of 
time series originating from well known dynamical systems, 
including continuous chaotic flows 
and discrete maps and to analyse their dependence on the details
of the phase space reconstruction.

\section{Acknowledgments}
The authors would like to thank Dr. G. Boulougouris, Prof. F. Diakonos and Prof.
D. J. Frantzeskakis for useful discussions 
and critical comments. E. P. acknowledges a PhD fellowship from the NCSR ``Demokritos". 
This research has been co-financed by the European Union (European Social Fund – ESF) 
and Greek national funds through the Operational Program ``Education and Lifelong Learning" 
of the National Strategic Reference Framework (NSRF) - Research Funding Program: THALES. 
Investing in knowledge society through the European Social Fund.


\end{document}